\documentclass[aip,jap, 12pt, preprint]{revtex4-1}
\usepackage{graphicx}
\begin{document}
\title{Proximity effect in Nb-Mo layered films:  Transition temperature and critical current dependence on period} 
\author{A.E. Walker} 
\altaffiliation{Univ. of Wisconsin, Madison, WI 53706}
\affiliation{Covenant College, Lookout Mountain, GA 30750}
\author{J. Veldhorst}
\affiliation{Covenant College, Lookout Mountain, GA 30750}
\author{D. M. Myers}
\affiliation{Covenant College, Lookout Mountain, GA 30750}
\author{Z. McElrath}
\altaffiliation{Skoodat LLC, Chattanooga, TN 37409}
\affiliation{Covenant College, Lookout Mountain, GA 30750}
\author{J.B. Lewis}
\altaffiliation{Washington University, St. Louis, MO 63130}
\affiliation{Covenant College, Lookout Mountain, GA 30750}
\author{P.R. Broussard}
\email{phill.broussard@covenant.edu}
\affiliation{Covenant College, Lookout Mountain, GA 30750}

\begin{abstract}
The behavior of the transition temperature and critical current density for a Mo/Nb repeated bilayer system  as a function of the number of periods was explored.  The measured values of the transition temperature are compared to the theoretical predictions for the proximity effect in the dirty limit.  We find that the transition temperature does not decrease as the number of periods increase.  In addition, inductive critical current density measurements also show a scaling that indicates the superconductivity properties are not dependent on the number of bilayers.
\end{abstract}

\maketitle

\section{Introduction}

The level of interest in the superconducting proximity effect for layered systems has continued to be very high with the advent of structures using ferromagnetic materials in proximity with superconductors directly or via normal metals\cite{Armenio, Alija, Char1}.  Theoretical studies of proximity systems continue to be carried out as predictions of the superconducting transition temperature of bilayers and trilayers are critical in their use as detectors, as well to understand the interactions of the layers.\cite{Martinis, Brammertz, Char2,Kushnir1}

The dependence of the critical temperature as the number of periods in a proximity system is varied has also been recently considered.\cite{Kushnir2}  In this work, the system N[SN]$^{m}$ was studied in the Nb/Cu system, showing how the transition temperature varied as the number of periods, $m$, were increased.  Here N refers to a normal metal layer and S to a superconducting layer.  An open question that still remains in light of this extensive work is exactly how does the critical temperature vary in the system [NS]$^{m}$?  The open question of how a system would transition from a single bilayer to a multilayer is still not fully understood, as discussed under the DeGennes-Werthamer (DGW) model,\cite{PRB1} which is equivalent to the single mode Usadel model.  In the work by Broussard,\cite{PRB1, PRB2} the implication is that the transition temperature of a [NS]$^{m}$ system would not change as $m$ is varied, even though it would change for both a N[SN]$^{m}$ as well as a S[NS]$^{m}$ system.   In a study of the Nb/Pd multilayer system, this result was used to calculate the transition temperature successfully, but no dependence on $m$ was studied.\cite{Falco}

In an attempt to bring some clarity to this question, we have carried out a study of the Mo/Nb layered system, growing the [NS]$^{m}$ system with $m$ varying from 1 to 4.  The Mo/Nb multilayered system has not been previously studied, although Mo/V has been studied extensively by Karkut {\it et al.}\cite{Karkut}  In that work, anomalies in the superconducting transition temperature were observed and partially linked to the properties of the vanadium layers.  In this work we have chosen the Mo/Nb system partly due to it not being studied before and the ability to see if what was observed for the Nb/Zr system\cite{PRB1}
would be replicated in a different system.  In addition we have carried out measurements of the inductive critical current (current in plane) as a function of temperature to see how this varies as the number of periods is changed, since as pointed out in the work on critical currents in Nb/Fe bilayers by Geers {\it et al.}\cite{Geers} that while transition temperatures probe the maximum value of the order parameter, critical currents involve an averaging over the entire layer thickness.

\section{Theoretical Considerations}

Initially, it was believed that increasing the number of periods in a [NS]$^{m}$ system would eventually be equivalent to an infinite multilayer,\cite{Gallagher} which would imply a reduction in  T$_{c}$ as $m$ increases, since an infinite multilayer is equivalent to a bilayer with layer thicknesses reduced by 1/2. Other work\cite{PRB1} however implies that the superconducting transition temperature (T$_c$) for the [NS]$^{m}$ layered system should not change as $m$ increases.  While it may initially seem counterintuitive that boundaries would be significant when approaching an infinite number of layers, viewing the system as analogous to a one-dimensional quantum potential well does lead us to expect this behavior.\cite{PRB2} In this analogy, each nonsuperconducting layer N is treated as a potential barrier and each superconducting layer S as a zero potential region. The amplitude of the wavefunction is related to the density of the superconducting pairs in the multilayer.

To compare to our measurements, we use the dirty limit DeGennes-Werthamer theory used in previous studies of bilayers.\cite{DeGennes, Werthamer} We use the formulation as laid out in previous work.\cite{PRB1}  Calculations can easily be carried out for our system, with all parameters measured except for the electronic specific heat coefficients.  For these we use published values of 7.8 and 2.0 mJ/(mole-K$^{2}$) for Nb and Mo, respectively.\cite{Kittel}

\begin{figure}
\begin{center}
\includegraphics*[width=6in]{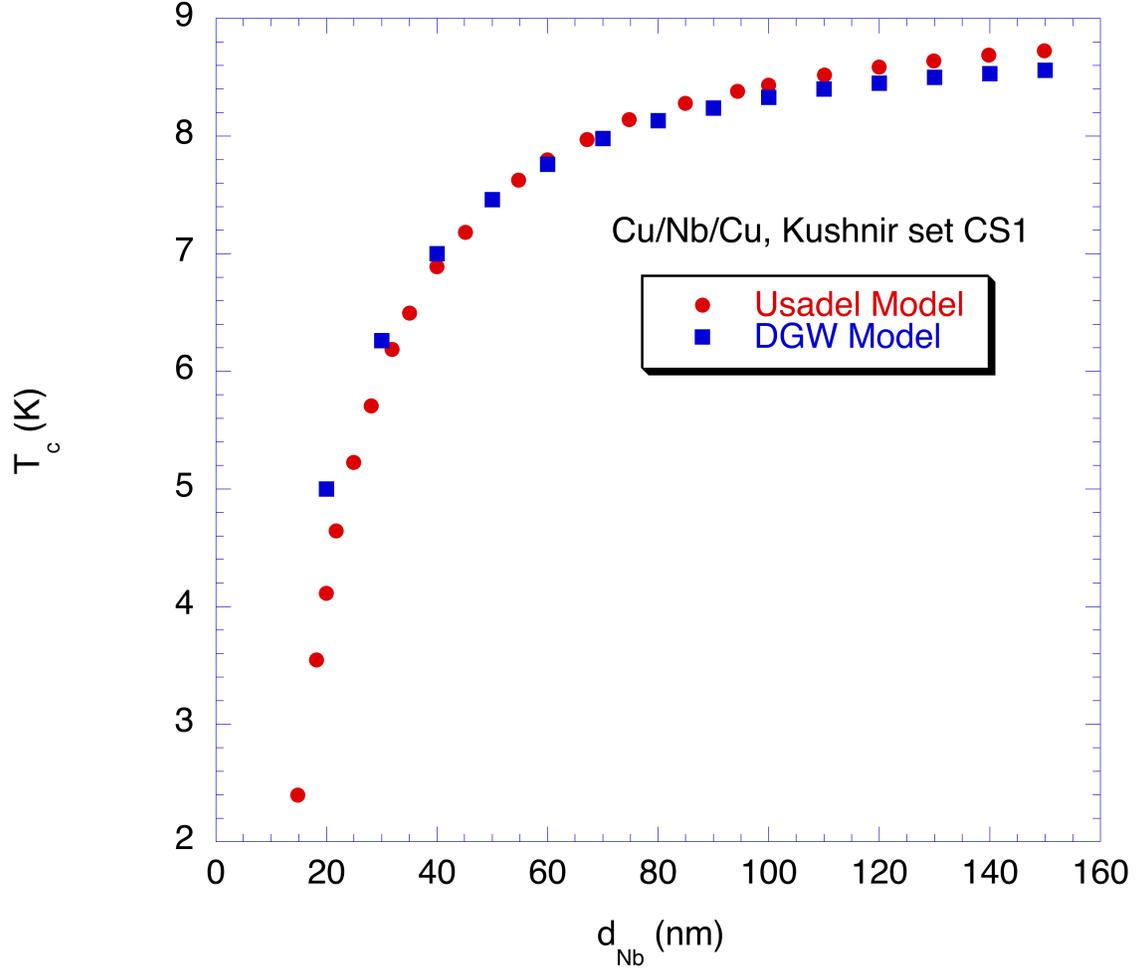}
\end{center}
\caption{Comparison between a DeGennes-Werthamer model and the Usadel model for the Cu/Nb/Cu trilayer set CS1 in Kushnir {\it et al.}\cite{Kushnir1}.  For the DGW model, the resistivity of the layers was taken from Kushnir {\it et al.} and the electronic specific heat coefficients were taken from Kittel.\cite{Kittel}}
\label{DGW}
\end{figure}

As mentioned earlier, there has been recent work\cite{Kushnir2} on how the T$_c$ will vary for a N[SN]$^{m}$ using the Usadel theory.  For this configuration of layers, the Usadel model predicts that T$_c$ increases as $m$ increases, which is what was also seen in the earlier modeling under DeGennes-Werthamer theory for a N[SN]$^{m}$ system.\cite{PRB1}  Since the Usadel model in the single mode limit and the DGW theory are the same, it would be important to know for our films if a single mode limit is appropriate.  Comparing the simple DGW model to the more complex Usadel predictions for the data of Kushnir {\it et al.}\cite{Kushnir1} in Fig. \ref{DGW} we find that the two agree within 0.2K for the niobium thickness range we are using.  In general the two sets agree over a wide range of niobium thickness. The lack of exact agreement at the extremes of the plots are due mainly to the DGW model having fewer fit parameters.  For the calculation shown in Fig. \ref{DGW}, there are no adjustable terms, as all have come either from the measured values in the paper by Kushnir {\it et al.} or in the case of the electronic specific heats, from Kittel.\cite{Kittel}  Considering this limitation, the agreement is encouraging.  Thus we feel that using the DGW model here, as it can be extended to multiple layers, is justified.  The Usadel model, as developed for multiple periods in a N[SN]$^{m}$ system depends on a center of symmetry,\cite{Kushnir2} which the [NS]$^{m}$ system does not have.

\section{Sample Preparation}

Our multilayers were prepared using DC magnetron sputtering in a high vacuum system. The background pressure, $P_b$, of the sputtering chamber was consistently less than 27$\mu$Pa before deposition.  Silicon substrates were clamped to a room temperature platform, which rotated at 10 rpm to ensure a uniform sample deposition distribution onto
 the substrates. Ultra high purity  (99.999\%)
Argon was used as the sputtering gas, with a pressure of  0.27 Pa during sputtering.  Separate DC power supplies operating at 250 W constant power were used for the Nb (S) and Mo (N) sources, with each sputter target of material rated at 99.95\% purity. Each source had a computer controlled mechanical shutter which could be closed to prevent the source from being deposited. The deposition rates for Mo and Nb were approximately 0.22 and 0.16 nm/s, respectively. Using these rates, we grew  36.9 nm Mo and 43.2 nm Nb layers.     Samples with 1-4 repeated bilayers were grown in different depositions.

\section{X-ray characterization}
\begin{figure}
\begin{center}
\includegraphics*[width=6in]{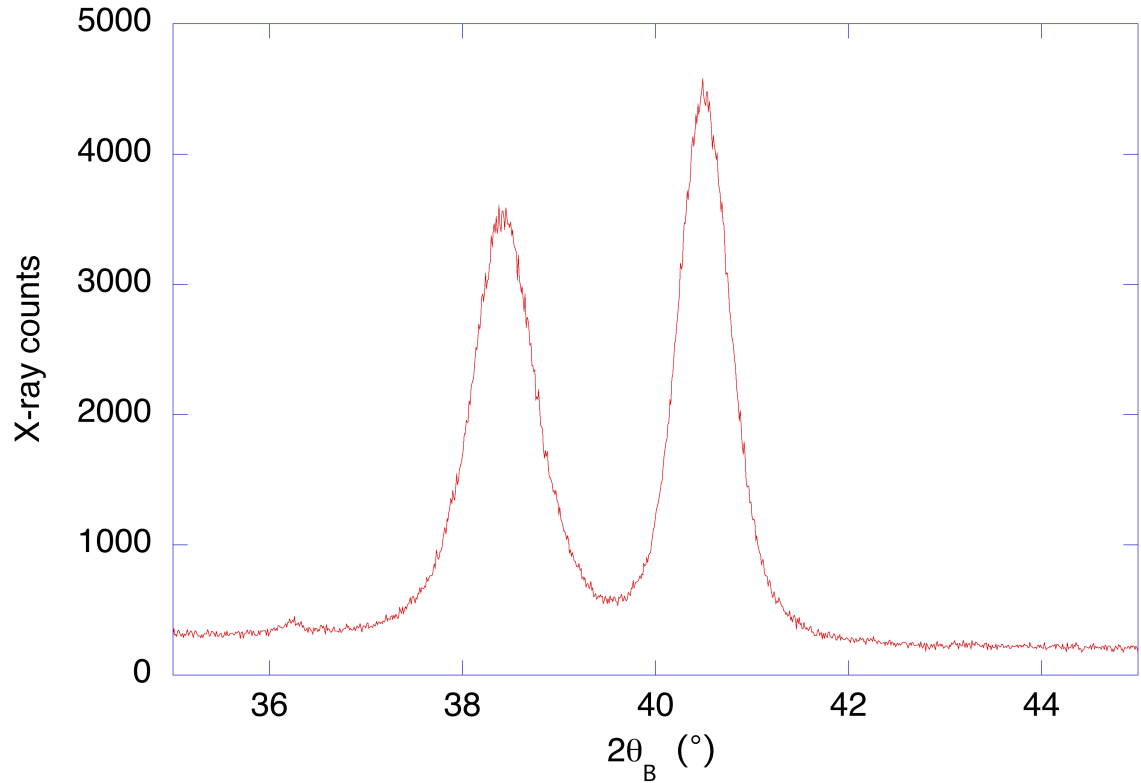}
\end{center}
\caption{X-ray diffraction along the sample normal for a two period (Mo/Nb) layered sample showing the (110) niobium and molybdenum lines.}
\label{XRD}
\end{figure}

In order to characterize the lattice structure and spacing of our samples X-ray diffraction data along the sample normal was collected, for individual layers and the two period bilayers with an example shown in Fig. \ref{XRD}.  The samples appear to have (110) texture, as is expected for samples of Nb and Mo grown at room temperature.  Gaussian curves were fit to the diffraction peaks to give the Bragg angle and the width of the diffraction peak.  The planar spacing ($d$) and Debye-Scherrer length ($t$) were then extracted according to the equations 
\[d = \frac{\lambda}{2sin(\theta_{B})}\]
and
\[t = \frac{0.9\lambda}{cos(\theta_{B})w},\]
where $\lambda$ is the wavelength of the X-ray beam, equal to 0.1542 nm, $\theta_{B}$ is the Bragg angle, and $w$ is the full width at half maximum.   The $d_{(110)}$ was 0.2343 nm for Nb and 0.2229 nm for Mo, 
 which points to the niobium layer being slightly stressed, while the molybdenum layer is  nearly at the bulk value.  The Debye-Scherer lengths are approximately 17 nm and 24 nm for niobium and molybdenum, respectively, both less than the layer thickness, but we notice that as the molybdenum layer is under less stress, the correlation length is also longer.  We see no evidence of multilayer peaks at these large values of layer thicknesses.
 
\section{Transport Measurements and Superconducting Transition Temperature}
To measure the transition temperature of each of the films, we first measured the electrical resistivity of the samples using the van der Pauw technique.\cite{VdP} Contacts were made with pressed indium on the perimeter of the sample, which was then pressed into an indium foil sheet against the OFHC (Oxygen Free High Conductivity) Cu sample stage.  Cooling was provided by a closed cycle cryocooler capable of reaching 6 K. The resistance data was gathered as the sample cooled from room temperature.  Once the material went superconducting, we warmed and cooled it around the transition temperature to establish consistency.  These measurements at low temperature were repeated for both of the van der Pauw configurations.  The temperature at which the transition occurred is taken to be the midpoint of the transition; the width of the transition (from 10\% to 90\% of the normal state resistivity) is typically 0.01K.  

Pure films of Nb and Mo of approximately 90 nm were grown and showed low residual resistivity ratios (RRR), indicating the samples are in the dirty limit of superconductivity.  For Nb the RRR was 3.0 and low temperature resistivity($\rho_{10\mathrm{K}}$) was 69 n$\Omega$-m, while for the Mo film RRR was about 1.7 and $\rho_{10\mathrm{K}}$ was 56 n$\Omega$-m.  The $T_{c}$ of our pure Nb film was 8.77 K.  We were unable to tell if the Mo film was superconducting, so we will consider it to be a normal metal for our purposes.(It is known that the $T_{c}$ of Mo is very sensitive to contamination and easily suppressed below the bulk value of 0.9 K.)

The electrical resistivity for the Mo/Nb layered films showed a RRR between that of Nb and Mo, at a value of 2.7 for the 4 period sample decreasing to 2.3 for the one period sample.   In an ideal situation, we would not expect RRR to change, but we did see both it and the thermal component
of the resistivity ($\rho_{\mathrm{290 K}}-\rho_{\mathrm{10 K}}$) changing, the latter increasing as the number of periods decreased.  The low temperature resistivity of the layered structures was of order 50-80 n$\Omega$-m.

\begin{figure}
\begin{center}
\includegraphics*[width=6in]{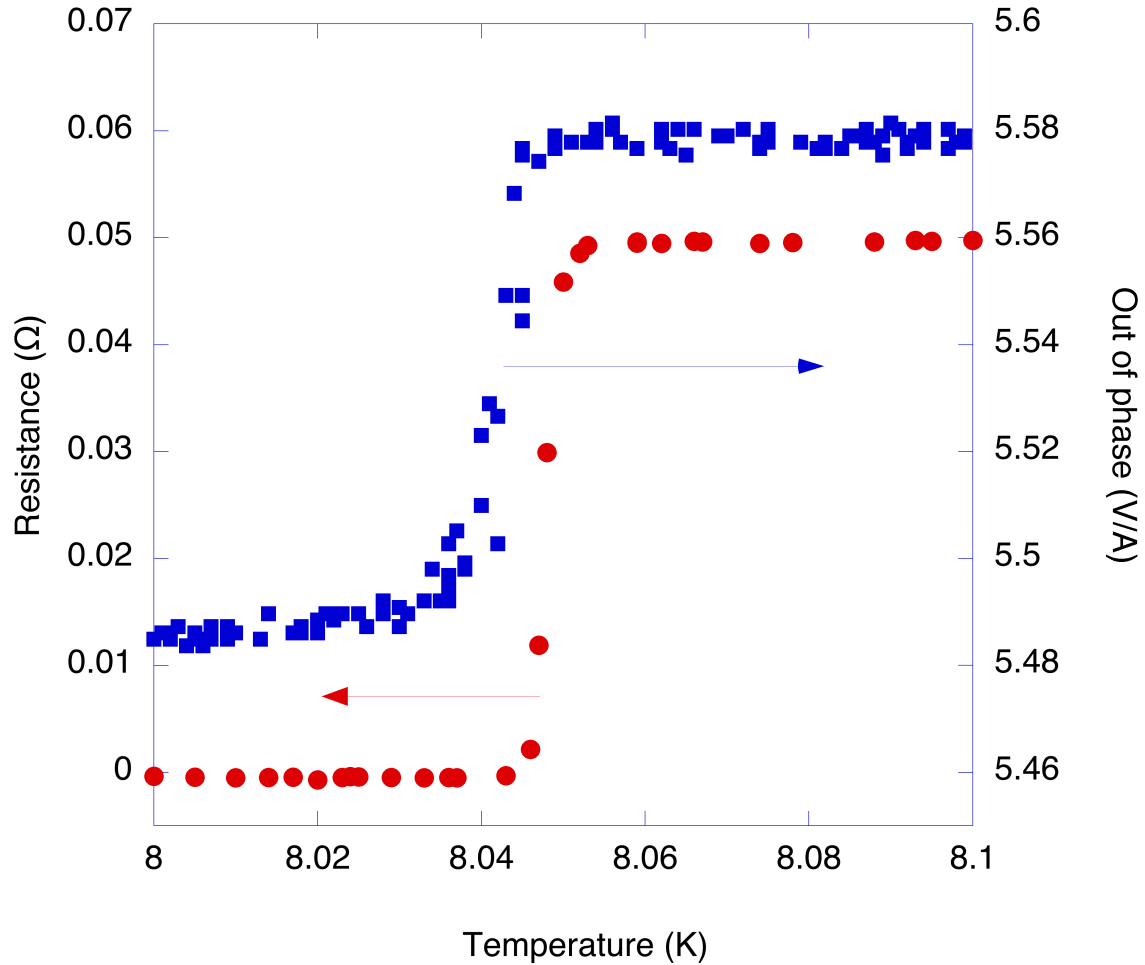}
\end{center}
\caption{Comparison of the superconducting transition for a 3 period Mo/Nb sample.  The resistive transition is shown by the filled circles
and the inductive transition is shown by the filled squares.}
\label{TcComp}
\end{figure}

To verify the quality of the samples, we also measured the superconducting transition using an inductive coil technique.\cite{Claassen}  A coil with 64 turns of 1/4 mm diameter wire with coil inner diameter 3.0 mm and 3.5 mm outer diameter was pressed against the film surface, and excited with a 100 kHz drive current of 0.1 mA rms.  The out of phase signal was measured using a lock-in method.  In Fig. \ref{TcComp} we show the comparison of the transitions for the resistive and inductive measurements and see as expected that the inductive transition starts as the resistive transition is ending.  The inductive transitions are also typically 0.01 K wide.

\section{Dependence on period}

\begin{figure}
\begin{center}
\includegraphics*[width=6in]{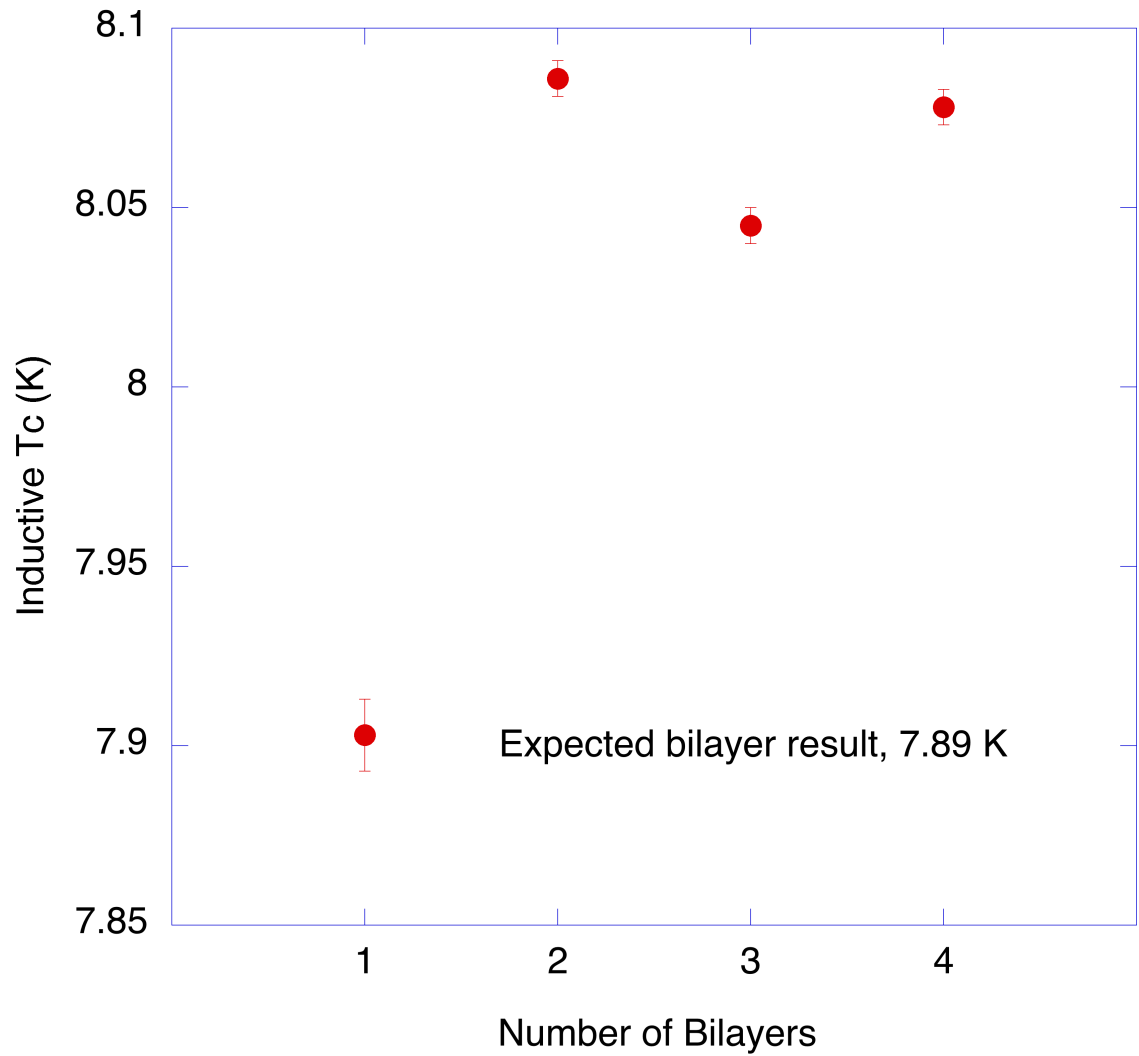}
\end{center}
\caption{Measured resistive superconducting transition temperatures
 for (Mo/Nb)$_{m}$ sequences vs. $m$ the number of bilayers.  In addition, the theoretical prediction for a bilayer is displayed.  The value for an infinite multilayer is much lower, at 6.91 K for this system.}
\label{TcPer}
\end{figure}

As stated before,\cite{PRB1} what has been expected for a [NS]$^{m}$ system as the number of periods ($m$) gets larger is for the system to go from the single to infinite case, while the theoretical investigation implies that the DeGennes-Werthamer theory would give
an unchanging transition temperature as the number of periods increases.  In Fig. \ref{TcPer} we show the resistive transition temperature for the Mo/Nb layered sequences, where the number of bilayers goes from 1 to 4.  Instead of the transition temperature decreasing as $m$ increases, we see what looks like an increase going from m=1 to m =2 with the values holding constant from then on.  The T$_{c}$ for the $m$=1 agrees very well with the prediction from the DGW model.  The value for the infinite multilayer result would be about 1 K lower, which is off the graph.  Clearly we do not see any decrease in the transition temperature towards the infinite multilayer result, which is very similar to what was observed in 1-3 period Nb/Zr layered samples\cite{PRB1} and continues to match those predictions.  The increase seen in this figure for $m > 1$ is not similar to the work by Kushnir {\it et al.}, who predicted a significant increase in T$_{c}$ ($\approx$ 0.5 K) as $m$ increased from 1 to 2.  In that work, since the fraction of superconductor is increasing with $m$, there is a large and sustained increase in T$_{c}$ as $m$ increases.  We only see the jump up at $m$ = 2.  This could simply be explained by the niobium layers improving their T$_{c}$ (due to both surfaces being protected from oxidation) by merely 0.15 K.   Since the transition temperatures remain nearly constant for $m\ge2$, we feel this is more in agreement with the [NS]$^{m}$ system.\cite{PRB1}

\section{Inductive Critical Currents}

\begin{figure}
\begin{center}
\includegraphics[width=5in]{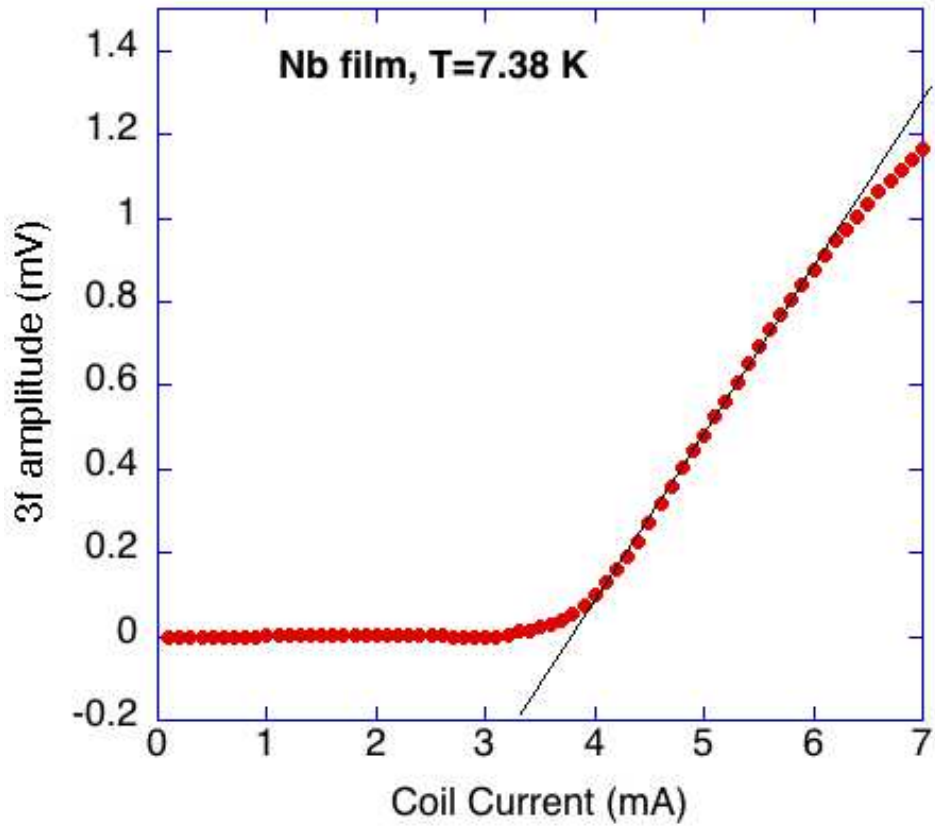}
\end{center}
\caption{3rd harmonic voltage vs. coil current for a pure Nb film at 7.38 K, showing how the critical current is defined}
\label{Icdef}
\end{figure}

To further probe the superconductivity of the multiple period structures, we measured the inductive critical currents in the films.\cite{Claassen}  We use a 300 turn coil (made of 25 $\mu$m wire with inner  and outer radial dimension of 1.05 mm and 1.95 mm, and an axial height of 0.46 mm) pressed against the samples, driven at 1 kHz and look for the nonlinear response at 3 kHz.  This obviously gives a current in plane type measurement.  The coil specifics and definition of the  critical current density are identical to that in Classeen {\it et al.}, as shown in Fig. \ref{Icdef}.
We point out that inductive critical current density measurements do not suffer from some of the issues that transport measurements do such as the pile up of current on the edges of the patterned film.\cite{Geers}  

\begin{figure}
\begin{center}
\includegraphics[width=5in]{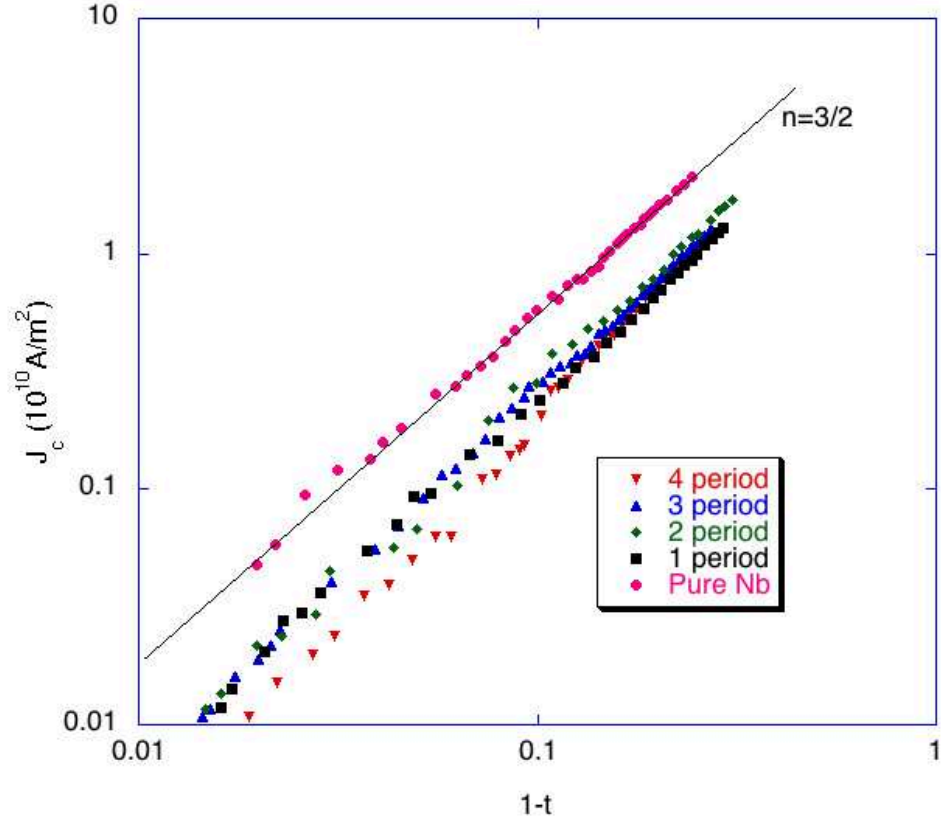}
\end{center}
\caption{Log-log plot of the inductive critical current density vs $(1-t)$ where $t$ is the reduced temperature for the pure niobium sample and the layered samples.  The line gives the expected Ginzburg-Landau dependence.}
\label{Jc}
\end{figure}

The measured critical current density as a function of temperature for the samples, shown in Fig. \ref{Jc}, follow a standard Ginzburg-Landau dependence, $J_{c}\propto(1-t)^{3/2}$ where $t$ is the reduced temperature, as expected.   Our pure niobium sample has a value of $J_{c}(0)$ that is an order of magnitude lower than predicted by the GL depairing formula in Geers' paper.   This result might be due to vortex formation, since the induced current path in the film will be of the order of the width of the coil, which is much larger than the coherence length.  What is far more relevant to our argument here is that the while the layered samples have lower critical current density compared to the pure niobium sample, the critical current densities for the layered samples collapse to essentially the same curve.   As stated in Geers, since critical current density reflects an average over the layers, then this average is also unchanging as the number of periods varies.  This gives even greater weight to the point that the order parameter shape is somehow remaining, on average, unchanged as the number of periods increases, which again goes against the usual perspective on how one would interpret the transition to an infinite multilayer result.

\section{Conclusions}

As has been seen before, the superconducting transition temperature for a repeating bilayer structure composed of normal and superconducting layers seems to remain constant as the number of periods goes from 1-4, which is in agreement with previous theoretical analysis but is non-intuitive.  The inductive critical current density is also seen to not depend on the number of periods.  Obviously the behavior for larger $m$ remains to be studied as well as other geometries, but this indicates work still needs to be done to fully understand the behavior of such systems with free boundary conditions.

\section{Acknowledgements}

The films were grown and measured on equipment funded by the
National Science Foundation under Grant DMR-0820025.  We gratefully acknowledge the help of Dr. C.B. Eom at University of Wisconsin for the X-ray diffraction data.   Film thicknesses were measured at the Georgia Tech Nanotechnology Research Center, a member of the National Nanotechnology Infrastructure Network, which is supported by the National Science Foundation (Grant ECS-0335765).

\end{document}